\documentclass[sigconf,natbib=true]{acmart}
\setcopyright{none}
\renewcommand\footnotetextcopyrightpermission[1]{} 
\newcommand{\name}{$\mathtt{AdCob}$}

\AtBeginDocument{%
  \providecommand\BibTeX{{%
    \normalfont B\kern-0.5em{\scshape i\kern-0.25em b}\kern-0.8em\TeX}}}

\setcopyright{acmcopyright}
\copyrightyear{2022}
\acmYear{2022}
\acmDOI{XXXXXXX.XXXXXXX}

\acmConference[Conference acronym 'XX]{Make sure to enter the correct
  conference title from your rights confirmation email}{June 03--05,
  2018}{Woodstock, NY}
\acmPrice{15.00}
\acmISBN{978-1-4503-XXXX-X/18/06}

\usepackage{algorithm}
\usepackage{algorithmic}
\usepackage{hyphenat}
\begin{document}

\title{Cross-channel Budget Coordination for \\Online Advertising System}

%
\author{Guangyuan Shen}
\authornote{Both authors contributed equally to this research.}
\author{Shenjie Sun}
\authornotemark[1]
\email{{shenguangyuan.sgy, shengjie.ssj}@alibaba-inc.com}
\affiliation{%
  \institution{Alibaba Group}
  \city{Hangzhou}
  \state{Zhejiang}
  \country{China}
}

\author{Dehong Gao}
\authornote{Corresponding Author}
\author{Shaolei Li}
\author{Libin Yang}
\email{gaodehong_polyu@163.com}
\affiliation{%
  \institution{Alibaba Group}
 \city{Hangzhou}
  \state{Zhejiang}
  \country{China}}

\author{Yongping Shi}
\author{Wei Ning}
\email{{yongping.syp, wei.ningw}@alibaba-inc.com}
\affiliation{
  \institution{Alibaba Group}\country{ }
}




\begin{abstract}
In online advertising (Ad), advertisers are always eager to know how to globally optimize their budget allocation strategies across different channels for more conversions such as orders, payments, etc. 
Ignoring competition among different advertisers causes objective inconsistency, that is, a single advertiser locally optimizes the conversions only based on its own historical statistics, which is far behind the global conversions maximization.
In this paper, we present a cross-channel \underline{\textbf{Ad}}vertising 
\underline{\textbf{Co}}ordinated \underline{\textbf{b}}udget allocation framework (\name) to globally optimize the budget allocation strategy for overall conversions maximization.
We are the first to provide deep insight into modeling the competition among different advertisers in cross-channel budget allocation problems.
The proposed iterative algorithm combined with entropy constraint is fast to converge and easy to implement in large-scale online Ad systems.
Both results from offline experiments and online A/B budget bucketing experiments demonstrate the effectiveness of \name.
\end{abstract}




\received{20 February 2007}
\received[revised]{12 March 2009}
\received[accepted]{5 June 2009}


\maketitle

\begin{figure*}[h]
  \includegraphics[width=\linewidth]{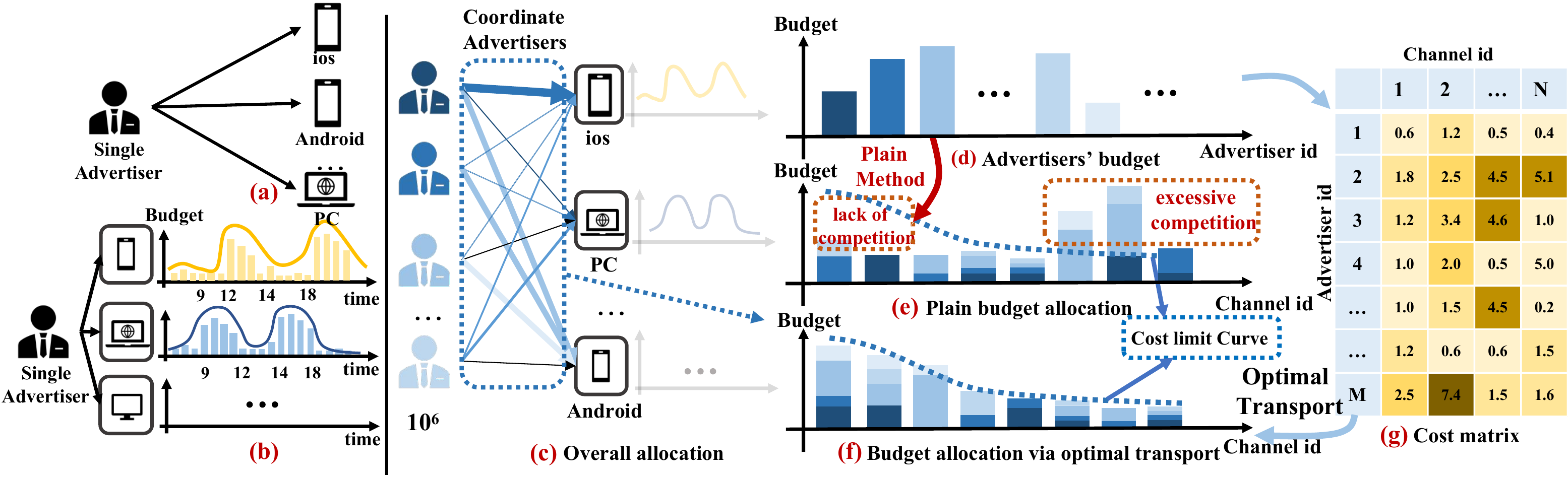}
  \caption{The framework of cross-channel budget allocation for online advertising. (a) and (b) show the cross-channel budget allocation for a single advertiser~\cite{zhao2019unified,li2018efficient,nuara2019dealing,nuara2018combinatorial}, while (c) shows the advertisers' interaction considered strategy. Base on the cost matrix~(g), we transfer advertisers' budgets to different channels with minimal conversion cost by optimal transport.}
   \label{fig:Diff}
\end{figure*}

\section{Introduction}
Budget management is essential for e-commerce search \underline{\textbf{Ad}}vertising (Ad) systems~\cite{balseiro2017budget,celli2022parity}, which simultaneously and deeply impacts the revenues of platforms and the performances of the advertisers~\cite{qin2015sponsored}.
As shown in Fig.~\ref{fig:Diff}(a), the overall traffic is composed of different channels according to the Ad display-ends such as mobile, laptop, etc.
%
If advertisers do not heuristically distribute their budgets, the Ad systems will roughly apply a first-come-first-served strategy. 
Under this situation, a first-come channel with poor quality may preempt the budgets of late-come channels with better quality, which leads to a conversions decrease within limited budgets.
%
Advertisers are thus eager to know how to reasonably distribute their limited budgets across different channels to obtain more conversions.

There is indeed some research working on optimizing the cross-channel budget allocation strategy for a single advertiser~\cite{zhao2019unified,li2018efficient,nuara2018combinatorial,nuara2019dealing}.
They assume that all the other advertisers are stationary, that is, maintaining their original allocation strategy. 
%
These works focus on one advertiser and optimize one's own conversions only based on local historical statistics.
None of them provide insights into modeling the competition among different advertisers.
%
\textbf{Ignoring this competition causes objective inconsistency}, that is, the Ad system is in a state of imperfect information competition.
We argue that under this situation, the Ad system will converge to a local optimal point of an inconsistent objective.
The local optimal point can be arbitrarily different from the global objective depending upon the proportion of advertisers adopting a greedy allocation strategy.
Specifically, if most advertisers greedily allocate their budgets to a small number of channels that they prefer, this will lead to such channels being allocated excessive budgets which may be beyond the upper bound that these channels can handle, i.e., excessive competition (please refer to Fig. \ref{fig:Diff}(e)). 
Such excessive competition not only leads to increasing costs for advertisers who have won the exposure opportunities in such channels but also lose potential exposure chances in other channels. 

In this paper, we focus on the prosperity of the market, coordinating all advertisers (as shown in Fig. ~\ref{fig:Diff}(c)) to optimize the global budget strategy across different channels for overall conversion maximization while maintaining platform revenue as much as possible.
%
To solve this problem, we cast the overall budget allocation problem as an \underline{\textbf{O}}ptimal \underline{\textbf{T}}ransport (OT) problem~\cite{galichon2018optimal,santambrogio2015optimal,villani2009optimal} and propose \textbf{\underline{Ad}vertising \underline{Co}ordinated \underline{b}udget allocation (\name) } approach, which satisfies the constraints of the advertiser's budgets and the channels' cost upper limits at the same time\footnote{The offline simulation first estimates the cost upper limits that different channels can undertake in one budget period (please refer to Fig.~\ref{fig:Diff}(e,f) cost upper limit curve).}.
%
%
As shown in Fig.~\ref{fig:Diff}(d,e,f), based on the cost matrix\footnote{As shown in Fig.~\ref{fig:Diff}(g), the averaged cost per conversion of a campaign on a channel can be regarded as the cost matrix in the OT problem. }, \name "transfers" advertisers' budgets to different channels with minimal "conversion cost" under bilateral constraints, which is more in line with the global optimal objective, i.e., global conversion maximization.
Almost all the advertisers can enjoy a better utility as the unreasonable competition (both excess and lack) is mitigated (Please refer to Fig. ~\ref{fig:Diff}(f)).
%
With a controllable entropy variable, \name~largely enhances objective consistency while preserving high search efficiency by searching space reduction~\cite{cuturi2013sinkhorn,peyre2017computational}.

Our main contributions are summarized below:
\begin{itemize}
    \item 
    We propose the global cross-channel budget allocation and cast it as Optimal Transport, which provides the first insight into modeling the competition among different advertisers and coordinating their budgets.
    \item We employ the iterative algorithm with entropy constraint which accelerates the training convergence and ensures large-scale implementation. Thanks to its simple framework, \name can be easily deployed to other Ad systems with cross-channel budget management needs.
    \item We have deployed the \name~framework in an online advertising system. The results from the offline experiments and the online A/B budget-bucketing experiments demonstrate the effectiveness of our proposed approach.  
    
\end{itemize}

\section{Related Work}
A recent strand of literature has considered different aspects of budget management in cross-channel Ad auctions.
The main difference to our work is that these works focus on a single advertiser, which is different from our global advertiser coordination.

Earlier literature~\cite{zhao2019unified} introduces the \underline{\textbf{M}}ultiple-\underline{\textbf{C}}hoice \underline{\textbf{K}}napsack (MCK)~\cite{pisinger1995minimal,sinha1979multiple} model to solve the cross-channel budget allocation of one single advertiser.
Some researchers take traffic fluctuations caused by time into consideration, and they cast the time-considered allocation as a reinforcement learning-based MCK problem~\cite{nuara2018combinatorial,li2018efficient}.
On this basis, the interactions among sub-campaigns\footnote{The number of sub-campaigns always less than 10.} are modeled in the allocation model~\cite{nuara2019dealing}.
All these methods ignore the interactions among a tremendous number of advertisers, that is, only working under the assumption that all the other advertisers keep the static strategy. 
These methods seem not suitable for real online Ad systems where millions of users bid to show their ads.

Besides, pacing methods are another series of budget management focusing on how to allocate the budget over the time blocks of a channel~\cite{xu2015smart} or how to adjust the budget cost rate according to the budget usage~\cite{agarwal2014budget}.
%
Pacing methods can also be regarded as cross-channel budget allocation by distributing the budgets across different time segmentations (channels).

\section{Method}
\subsection{Optimal Transport Problem}
Optimal transport (OT) is the problem of moving goods from one set of warehouses to another set of destinations while minimizing certain cost functions~\cite{peyre2017computational} (please refer to Eq.~\ref{plain_objective}).
For example, suppose that we have $N$ warehouses (located by $\left\{x_{i}\right\}_{i=1}^N$), the number of goods in each warehouse is $\{G_i\}_{i=1}^{N}$, and need to be moved to $M$ different places (located by $\{y_j\}_{j=1}^{M}$). The quantity of demanded goods of each destination is $\{D_j\}_{j=1}^{M} $, and the unit transportation cost between the $i^{th}$ warehouse and the $j^{th}$ destination constructs the cost matrix $\mathbf{C}, \{c(x_i,y_j)\}_{i,j=1}^{N,M}$.
Then, the OT problem can be formulated as follows,
\begin{align}
    L=&\underset{\Gamma}{\arg \min } \sum_{i=1}^{N} \sum_{j=1}^{M} \Gamma_{i j} c\left(x_i, y_j\right) \label{plain_objective}\\
    s.t.~ &\forall i\in \{1,..,N \}~~\sum\limits_j {{\Gamma _{ij}}} = {G_i} \nonumber \\
    &\forall j\in \{1,..,M \}~~\sum\limits_i {{\Gamma _{ij}}} = {D_j} \nonumber
\end{align}
where $\Gamma$ is the transport matrix to optimize, and $\Gamma_{ij}$ denotes the number of goods sent from the $i^{th}$ warehouse to the $j^{th}$ destination. 
Moreover, we must have $\sum_{i} G_{i} = \sum_{j} D_{j}$, since the total quantity of goods will not change.
If we are in the \textbf{unbalance} situation where $\sum_{i} G_{i} \leq \sum_{j} D_{j}$, we can set a virtual warehouse (located in $x_{i+1}$) with $|\sum_{i} G_{i}-\sum_{j} D_{j}|$ goods stored in. 
When we set the cost between the $(N+1)^{th}$ warehouse and the $j^{th}$ destination equal to 0, i.e., $\forall j, c(x_{i+1,j})=0$, we can convert the unbalance OT to balance.

\subsection{Budget Allocation via Optimal Transport}
In an online Ad system, advertisers are allowed to create Ad campaigns and the budget allocation in this paper refers to the budget allocation of each campaign.
We reformulate the global cross-channel budget allocation as an unbalanced OT.
Campaigns' budgets are viewed as the goods in warehouses while the channel cost upper limits are viewed as the demanded goods at each destination. 

Suppose that we have $N$ Ad campaigns with budgets $\boldsymbol{b}:=\{b_{i}\}_{i=1}^N$, and have $M$ channels with different daily cost upper limits $\boldsymbol{h}:=\{h_{j}\}_{j=1}^M$ (the cost upper limit estimation refers to Sec.~\ref{sec:cost limit}).
%
%
%
We try to maximize the number of conversions by optimizing the budget allocation matrix $\mathbf{P}:=\{\mathbf{P}_{i,j}\}_{i=1,j=1}^{N,M}$, where $\mathbf{P}_{i,j}$ denotes the budget that the $i^{th}$ campaign distributes to the $j^{th}$ channel.
When the total budget is fixed, the objective converts to minimize the global CPC (\underline{\textbf{C}}ost \underline{\textbf{P}}er \underline{\textbf{C}}onversion), i.e., minimize the linear combination of different $\mathrm{CPC}_{i,j}$ with weight $\mathbf{P}_{i,j}$.
Here $\mathbf{C}:=\mathrm{\{CPC_{\mathit{i,j} }\}}_{i=1,j=1}^{N,M}$ denotes the CPC of the $i^{th}$ campaign on the $j^{th}$ channel (please refer to Eq.~\ref{budget_objective}).
Faced with the sparse nature of Ad conversion action, the calculation of the cost matrix is difficult (for more details, please refer to Sec.~\ref{sec:cost matrix}).
Moreover, in practice, the sum of the cost upper limits of all channels is always greater than the sum of all the budgets. 
We can make up a virtual campaign with virtual budget $b_{N+1}:=|\sum_{i}b_{i} -\sum_{j}h_{j}|$ to bridge the budget gap and simply set the CPC of this campaign on each channel as 0, i.e., $\forall~j,~\mathbf{C}_{N+1,j}:=\mathrm{CPC}_{N+1,j} = 0$.
The formal problem formulation is as follows:
\begin{align}
    L=&\underset{\mathbf{P}}{\arg \min } \sum_{i=1}^{N} \sum_{j=1}^{M}\mathbf{P}_{i,j} \mathbf{C}_{i,j} \label{budget_objective}\\
    s.t.~ &\forall i\in \{1,..,N+1 \}~~\sum\limits_j {{\mathbf{P}_{i,j}}} = {b_i} \nonumber \\
    &\forall j\in \{1,..M \}~~\sum\limits_i {{\mathbf{P}_{i,j}}} = {h_j} \nonumber
\end{align}
Obviously, this is a large-scale linear programming problem with tremendous numbers of constraints. The complexity of the greedy solution is $\mathcal{O}(N^3logN)$~\cite{peyre2017computational}, the iteration speed is too slow to meet the model iteration requirements of large-scale Ad scenarios.

\subsection{Iterative Solution with Entropy Constraint}
The problem in Eq.~\ref{budget_objective} is a special linear programming problem, and advanced linear programming solving algorithms can be used to solve it.
However, when faced with such large-scale problems in online Ad, the advanced algorithm based on the interior point method~\cite{nemirovski2004interior} still has great limitations~\cite{peyre2017computational}.
Some researchers served that, the problem above can be solved in a practical and scalable way by adding an entropy penalty and using the matrix scaling Sinkhorn algorithm~\cite{cuturi2013sinkhorn}. 
The new objective of the problem is
\begin{equation}
    \label{Equation:sinkhorn}
      L=\underset{\mathbf{P}}{\arg \min }\sum_{i} \sum_{j}\mathbf{P}_{i,j} \mathbf{C}_{i,j} - \epsilon \mathbf{H}(\mathbf{P}),
\end{equation}
where entropy $\mathbf{H}(\mathbf{P}) := -\sum_{i} \sum_{j} \mathbf{P}_{i,j}(\mathrm{log}(\mathbf{P}_{i,j})-1)$, and $\epsilon$ is the coefficient of entropy regularization.
Since the objective in Eq.~\ref{Equation:sinkhorn} is an $\epsilon$-strongly convex function, it has a unique optimal solution~\cite{peyre2017computational}.
Introducing two dual variables $\mathbf{f} \in \mathbb{R}^N+1, \mathbf{g} \in \mathbb{R}^M$ for each marginal constraint, the Lagrangian of Eq.~\ref{Equation:sinkhorn} reads
\begin{equation}
    \mathcal{E}(\mathbf{P}, \mathbf{f}, \mathbf{g})=\langle\mathbf{P}, \mathbf{C}\rangle-\varepsilon \mathbf{H}(\mathbf{P})-\left\langle\mathbf{f}, \mathbf{P} \mathbf{1}_{N+1}-\mathbf{P}\right\rangle-\left\langle\mathbf{g}, \mathbf{P}^{\mathrm{T}} \mathbf{1}_M-\mathbf{h}\right\rangle,
\end{equation}
where $\langle~,~\rangle$ denotes Frobenius dot-product.
First-order conditions yield
\begin{equation}
    \frac{\partial \mathcal{E}(\mathbf{P}, \mathbf{f}, \mathbf{g})}{\partial \mathbf{P}_{i, j}}=\mathbf{C}_{i, j}+\varepsilon \log \left(\mathbf{P}_{i, j}\right)-\mathbf{f}_i-\mathbf{g}_j=0,
\end{equation}
which result, for an optimal $\mathbf{P}$ coupling to the regularized problem, in the expression $\mathbf{P}_{i, j}=e^{\mathbf{f}_i / \varepsilon} e^{-\mathbf{C}_{i, j} / \varepsilon} e^{\mathbf{g}_j / \varepsilon}$.
We iterate over $\{f_{l}\}$ and $\{g_{l}\}$ sequenced by the equations (a, b) in Algorithm~\ref{alg:allocate} until convergence. 
%
The $\{f_{l}\}$ and $\{g_{l}\}$ sequences essentially represent how the solution $\mathbf{P}$ budget allocation matrix satisfies the bilateral constraints.
We alternately satisfy the campaigns’ budget constraints and channel cost upper limit constraints by alternately iterating the sequences $\{f_{l}\}$ and $\{g_{l}\}$, respectively.

The $\epsilon$ controls the strength of the regularization. 
As the $\epsilon$ goes to zero, more accurate solutions can be obtained while the campaign’s budget will be centrally allocated to certain channels bringing numerical instability.
We present the auction results with different $\epsilon$ parameters in the experiment part, please refer to Sec.~\ref{Sec:Entropy} .
\begin{algorithm}[t]
    \caption{Sinkhorn iteration procedure}
    \label{alg:allocate}
    \begin{algorithmic}
        \STATE {Input:} Cost matrix $\mathbf{C}_{i,j}$, marginals $\mathbf{P} \in \mathbb{R}^{N+1}$, $\mathbf{h} \in \mathbb{R}^{M}$, entropy coefficient $\epsilon$
        \STATE {Initialize:} $\mathbf{f} = \mathbf{0}^{N+1}, \mathbf{g} = \mathbf{0}^{M+1}$
        \WHILE{no convergence}
            \STATE $\boldsymbol{f}_i=-\epsilon \log \left(\sum_j \exp \left(-\frac{C_{i j}-\boldsymbol{g}_j}{\epsilon}\right)\right)+\epsilon \log \boldsymbol{b}$\qquad\qquad\qquad\quad(a)
            \STATE $\boldsymbol{g}_j=-\epsilon \log \left(\sum_i \exp \left(-\frac{C_{i j}-\boldsymbol{f}_i}{\epsilon}\right)\right)+\epsilon \log \boldsymbol{h}$\qquad\qquad\qquad\quad(b)
            \STATE $\mathbf{P}_{i, j}=e^{\mathbf{f}_i / \varepsilon} e^{-\mathbf{C}_{i, j} / \varepsilon} e^{\mathbf{g}_j / \varepsilon}$
        \ENDWHILE
         \STATE {Output:} Feasible allocation matrix $\mathbf{P}_{i,j} \in \mathbb{R}^{(N+1) \times M}$
    \end{algorithmic}
\end{algorithm}
\section{Implementation Details}
\subsection{Estimated Cost Upper Limit}\label{sec:cost limit}
We use an offline simulated auction system~\cite{wei2019optimal} to estimate the cost upper limit of each channel.
By removing budget constraints for all Ad requests, all matching campaigns will be recalled as the impression candidate, and the bidding, uGSP auction~\cite{wilkens2017gsp,edelman2007internet,zhang2021optimizing} will be executed in order.
The average cost of each channel in the past 30 days is counted as the estimated cost upper limit of the channel.
\subsection{Estimate Cost Matrix}\label{sec:cost matrix}
%
For large-scale model deployment, we make statistics of the 30 days  CPC of a campaign on a channel as the cost $\mathbf{C}_{i,j}$ to construct the cost matrix $
    \mathbf{C}_{i,j} := \mathrm{CPC}_{i,j} = \frac{cost~30~days_{i,j}}{total~conversions~30~days_{i,j}}$.
%
%
%
%
In practice, we face two challenges and the corresponding solutions are
\begin{itemize}
    \item \textbf{Conversion actions are inherently sparse}, i.e, there are many campaign-channel pairs possessing no conversion action. We use a combination of estimated conversion rate and real conversion to count the number of conversions, so as to alleviate the sparsity of the conversion itself.
    \item \textbf{Partial cold start campaign}, i.e., some Ad campaigns have no cost on some channels. We use the average cost per conversion of the Ad campaign itself as its cost matrix
\end{itemize}

\section{Experiment}
\subsection{Offline Setting.}
We experimentally evaluate our cross-channel budget allocation method (\name) in an offline setting using a simulated auction system~\cite{wei2019optimal} and real-world datasets collected from our real online advertising system without any sampling.
\subsubsection{Baselines.}
Apart from the plainest first-come-first-served (FCFS) method, two other relevant budget allocation methods termed IDIL~\cite{nuara2019dealing} and unified budget allocation~\cite{zhao2019unified} have been introduced to our experiments. 
All these prior methods focus on only one advertiser, so we directly use these budget allocation methods for 40$\%$, 80$\%$ advertisers, regardless of whether these advertisers will generate unreasonable competition, resulting in a decline in platform revenue.
In addition, we did small treatments on them, for example:
\begin{itemize}
    \item IDIL~\cite{nuara2019dealing}: we ignore the strategy of allocating different budgets on different days since we focus on cross-channel budget allocation in this paper.
    \item Unified budget allocation~\cite{zhao2019unified}: We use a linear model to approximate the ROI curve.
\end{itemize}
\subsubsection{Data set.}
We evaluate our method with an advertising data set that was collected from a real-world Internet e-commerce company, where all advertises compete for more conversion, such as purchase behavior, and inquiry behavior. 
The real dataset covers nearly two hundred thousand campaigns.
The real dataset contains tens of millions of records with the following auction information:
\begin{itemize}
    \item \underline{\textbf{P}}redicted \underline{\textbf{C}}lick \underline{\textbf{T}}hrough \underline{\textbf{R}}ate (pCTR), \underline{\textbf{p}}redicted \underline{\textbf{C}}on\underline{\textbf{V}}ersion \underline{\textbf{R}}ate (pCVR) that describe user preferences for different items, predicted by Deep Interest Network~\cite{zhou2018deep}.
    \item Real bid price, generated by OCPC~\cite{zhu2017optimized} bidding strategy based on pCTR, pCVR, etc.
    \item Click action, click or not.
    \item Conversion action, conversion or not.
    \item Advertiser overall budget and remaining budget.
\end{itemize}
Each auction includes 5, 10, and 20 ad slots and 500 or 750 advertisers bidding for impressions.
Different channels have different numbers of ad slots.
In contrast, previous work only considers 5 slots and 10 advertisers bidding on a synthetic data set.
We will release the desensitized data set to help researchers better understand our method.
\subsubsection{Simulation system and metrics}
In our offline simulated auction system, we will traverse each traffic record block by block according to its timestamp. 
Each block contains all traffic records within fifteen minutes.
For each record, we implement a strict Generalised second-price auction~\cite{edelman2007internet}.
Before executing the auction, we rigorously check that the whether the recalled ad campaign has run out of its budget. If this ad campaign runs out of its budget, it will be offline immediately and will not participate in subsequent auctions.

The goal of global budget allocation is to maximize the overall conversions of winning impressions from all the Ad campaigns. 
Here we report the averaged \underline{\textbf{C}}ost \underline{\textbf{P}}er \underline{\textbf{C}}onversion (CPC), total \underline{\textbf{Conv}}ersions (Conv) and platform total \underline{\textbf{Rev}}enue (Rev) in a budget period\footnote{A budget period is generally 24 hours.}.
Because we cannot know the real click conversion behavior in the offline experiment, we will use the estimated display revenue of all ad campaigns that have received ad impressions as Revenue (Rev), and the sum of the conversion rates as the number of conversions (Conv).
In order to avoid the leakage of sensitive data, we normalize all the metrics, i.e., set the base method to 1.00 and report the percentage change when different budget strategies are turned on. 

\subsubsection{Offline results.}
\begin{figure}[ht]
  \includegraphics[width=\linewidth]{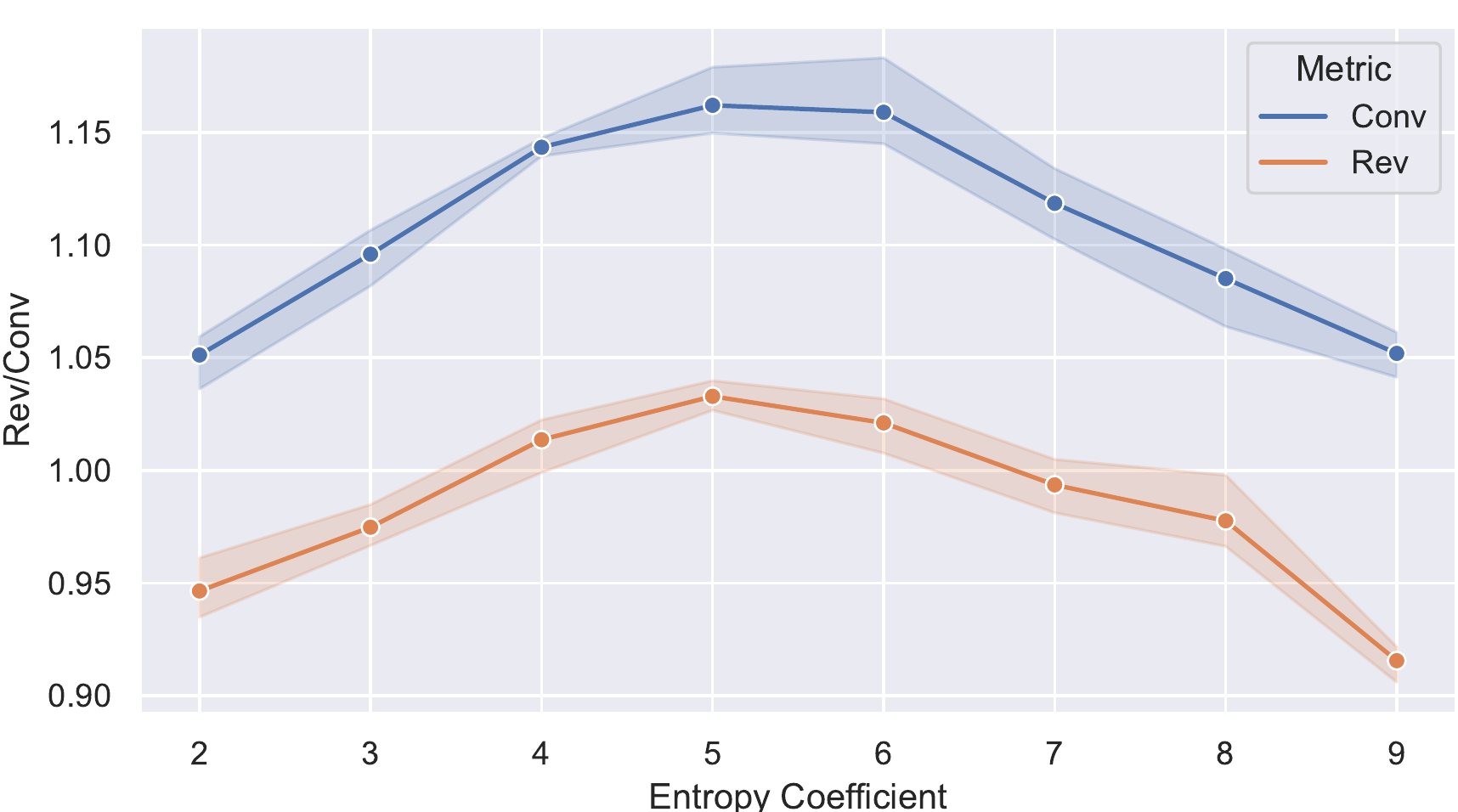}
  \caption{Rev and Conv of \name with different entropy $\epsilon$ in the offline experiment. As the $\epsilon$ increases, both the revenue and conversions first rise and then fall.}
   \label{fig:entropy}
\end{figure}

\begin{table*}[tb]
  \caption{According to the metrics (e.g., revenue, conversions, CPC), \name (with $\epsilon = 5.50$) achieves the best in the offline experiments compared to BASE, Greedy. Advertisers enjoy a 13.6 $\%$ CPC reduction and 19.1$\%$ conversion increase. At the same time, platform enjoys a 2.9$\%$ revenue increase. 
  }
  \begin{tabular}{cccl}
    \toprule
    Algorithm & Rev & Conv & CPC\\
    \midrule
    BASE & 1.000(---) &  1.000(---) & 1.000(---)\\
    40$\%$ IDIL~\cite{nuara2019dealing} & 0.981(-1.9$\%$)&0.947(-5.3$\%$)& 1.018(+1.8$\%$)\\
    80$\%$ IDIL~\cite{nuara2019dealing} & 0.973(-2.7$\%$)&0.901(-9.9$\%$)& 1.056(+5.6$\%$)\\
    40$\%$ Unified budget allocation~\cite{zhao2019unified} & 0.984(-1.6$\%$)&0.956(-4.4$\%$)& 1.016(+1.6$\%$)\\
    80$\%$ Unified budget allocation~\cite{zhao2019unified} & 0.965(-3.5$\%$)&0.910(-9.0$\%$)& 1.052(+5.2$\%$)\\
    \name(Ours) & 1.029(\textbf{+2.9}$\%$)&1.191(\textbf{+19.1$ \%$})& 0.864(\textbf{-13.6$\%$})\\
    \bottomrule
    \end{tabular}
    \label{tab:offline}
\end{table*}
We run extensive experiments on the real dataset to validate the effectiveness of the proposed approach.
Tab.~\ref{tab:offline} presents the offline benefit of revenue, conversions, and CPC by using our method.
Fig.~\ref{fig:entropy} captures the impact of the coefficient of entropy regularization $\epsilon$.
As the results show:

\textbf{CPC Reduction.}
The reason behind CPC reduction is that we \textbf{coordinate} all advertisers to optimize their conversion, fully considering the interaction among different advertisers, where we can effectively avoid excessive competition. Almost all advertisers can achieve more conversions within a prefixed budget.

\textbf{Revenue and Conversion Increase.}
In this paper, we focus on optimizing the conversions of all advertisers while maintaining the revenue of the platform.
The overall revenue has also increased even though we did not optimize revenue, as the budget utilization rate of some Ad campaigns has increased.
There are two kinds of ad campaigns existing in the ad platform, the former always has a relatively high bid and can easily run out of the budget, and the latter's bids are relatively low and may be in some unpopular track, barely able to spend their budget.
In the past, such Ad campaigns with relatively low bids cannot spend their budgets as there are unreasonable competition.
With the help of global budget management, the head campaigns (such campaigns with relatively high bids) will give priority to more suitable traffic channels, thus giving up some unsuitable channels.
The display opportunities on such "unsuitable" channels are obtained by the Ad campaigns in the middle and tail, improving their budget utilization rate.
The increase in conversions is caused by the overall CPC decreases and the budget utilization rate increase.

\textbf{Local Methods Cannot Work Well.}
When the proportion of advertisers who use local allocation~\cite{nuara2019dealing,zhao2019unified} in the entire Ad platform increases, the corresponding number of overall conversions and revenue will decrease.
This is because there is excessive competition in some channels, as too many advertisers greedily distribute their budgets based on their local historical data preferences. 

\textbf{The Impact of Entropy Coefficient $\epsilon$.}\label{Sec:Entropy}
In Fig.~\ref{fig:entropy}, we use different budget allocation matrix $\mathbf{P}_{i,j}$ that are computed by different $\epsilon$ to conduct auction simulations. 
As the $\epsilon$ gradually increases, the total number of conversions first increases and then decreases.
We analyze that this is because as the entropy term keeps getting bigger, the budget allocation becomes less sparse, but the solution (allocation strategy) also deviates more from the optimal solution.
Therefore, in our real online deployment, we will select the most appropriate $\epsilon$ based on the offline experiment results.
%
%
\subsection{Online Setting}
\subsubsection{Online Budget Bucketing A/B Test}
\begin{figure}[htb]
  \includegraphics[width=\linewidth]{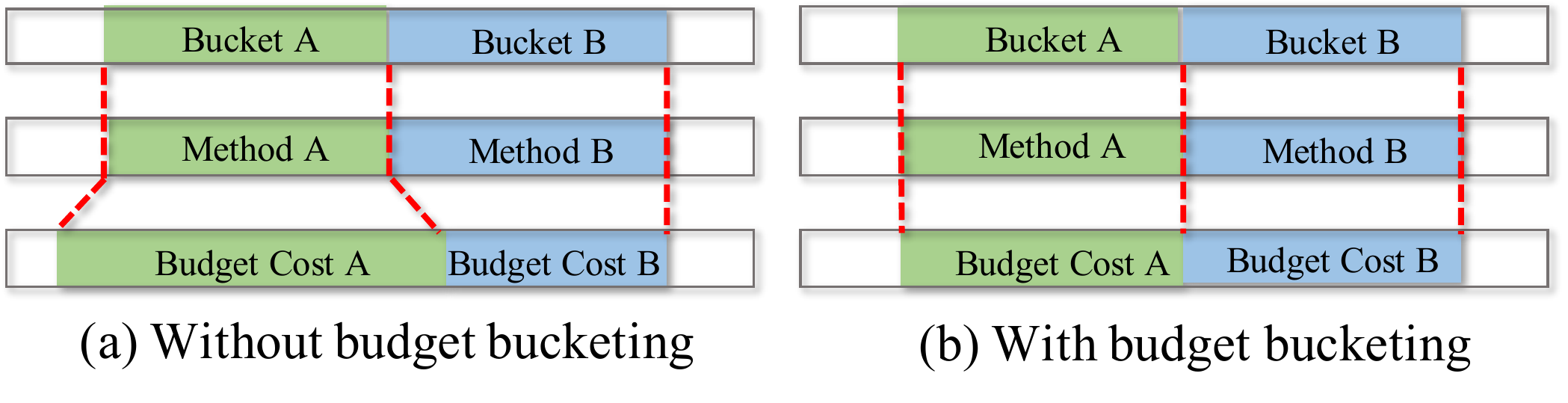}
  \caption{Visualization of Budget Bucketing. }
  \label{fig:bucket}
\end{figure}
In a typical A/B test, in order to compare control A with a treatment B, members are randomly split into two groups with one group receiving the control and the other group receiving the treatment. This approach is not directly applicable to what we wanted to test.

As shown in Fig.~\ref{fig:bucket}(a), when we perform budget management in bucket B but impose no budget restrictions in bucket A, bucket A will consume more budget than Budget B, i.e., getting a negative result. 
Here we introduce a simple yet effective experiment paradigm for correctly showing the impact of budget management. 
Learning from the work of other platform\footnote{https://support.google.com/searchads/answer/7518994?hl=en}, we will not only divide the whole traffic into two parts but also divide the total budget of a campaign into two parts.
As shown in Fig.~\ref{fig:bucket}(b), if a traffic hits bucket A, then it only uses the budget of bucket A without affecting that of bucket B, which ensures that the experimental and control bucket do not compete for the budget.
%
%
\subsubsection{Online results.}
Here we report the real results that we collect the traffic record from our online advertising system for 30 days. With OT budget allocation, we can help advertisers spend their budgets on high-quality conversion traffic while considering the interaction among different advertisers and avoiding excess competition on certain channels. To our best knowledge, we are the first to provide the online production experiment to validate the effectiveness of the proposed algorithm.
%
\begin{table}[ht]
  \caption{According to the numbers of online metrics (e.g., clicks, conversions, CPC), \name can enjoy a 24.6$\%$ conversion increase and a -17.6$\%$ CPC reduction. These results validate the effectiveness of our method.}
  \label{tab:online}
  \begin{tabular}{cccl}
    \toprule
    Algorithm & Click & Conv & CPC\\
    \midrule
    BASE & 1.000 & 1.000 & 1.000\\
    \name(Ours)&  0.968(\textbf{-3.2}$\%$) & 1.246(\textbf{+24.6$\%$}) & 0.824(\textbf{-17.6}$\%$)\\
  \bottomrule
\end{tabular}
\end{table}

\section{Discussion}
This paper presents a cross-channel budget management framework where we coordinate all competing advertisers to allocate a limited budget to different channels in order to maximize the overall conversion. In other words, we focus on market-making (which is very important for the ad platform) while maintaining or promoting platform revenue as much as possible.
In the future, we plan to present a more comprehensive theoretical analysis of the Nash equilibrium efficiency with game theory.
We also have an interest in combining the RL-based method to enhance our framework by dynamically adjusting the cost matrix.
As for the limitation, the method we propose is more suitable for advertisers who use auto-bidding techniques like OCPX~\cite{zhu2017optimized}. For those advertisers who bid independently, they might adjust their behavior (i.e., lower their offer or increase their offer) as a function of maximizing their own utility. We currently only apply our method on ad campaigns with automatic bidding, and we will work hard to be compatible with those who might adjust their behavior in the future.

\bibliographystyle{ACM-Reference-Format}
\bibliography{sample-base}


\begin{thebibliography}{23}


\ifx \showCODEN    \undefined \def \showCODEN     #1{\unskip}     \fi
\ifx \showDOI      \undefined \def \showDOI       #1{#1}\fi
\ifx \showISBNx    \undefined \def \showISBNx     #1{\unskip}     \fi
\ifx \showISBNxiii \undefined \def \showISBNxiii  #1{\unskip}     \fi
\ifx \showISSN     \undefined \def \showISSN      #1{\unskip}     \fi
\ifx \showLCCN     \undefined \def \showLCCN      #1{\unskip}     \fi
\ifx \shownote     \undefined \def \shownote      #1{#1}          \fi
\ifx \showarticletitle \undefined \def \showarticletitle #1{#1}   \fi
\ifx \showURL      \undefined \def \showURL       {\relax}        \fi
\providecommand\bibfield[2]{#2}
\providecommand\bibinfo[2]{#2}
\providecommand\natexlab[1]{#1}
\providecommand\showeprint[2][]{arXiv:#2}

\bibitem[Agarwal et~al\mbox{.}(2014)]%
        {agarwal2014budget}
\bibfield{author}{\bibinfo{person}{Deepak Agarwal}, \bibinfo{person}{Souvik
  Ghosh}, \bibinfo{person}{Kai Wei}, {and} \bibinfo{person}{Siyu You}.}
  \bibinfo{year}{2014}\natexlab{}.
\newblock \showarticletitle{Budget pacing for targeted online advertisements at
  linkedin}. In \bibinfo{booktitle}{\emph{SIGKDD}}. \bibinfo{publisher}{ACM},
  \bibinfo{address}{New York, USA}.
\newblock


\bibitem[Balseiro et~al\mbox{.}(2017)]%
        {balseiro2017budget}
\bibfield{author}{\bibinfo{person}{Santiago Balseiro}, \bibinfo{person}{Anthony
  Kim}, \bibinfo{person}{Mohammad Mahdian}, {and} \bibinfo{person}{Vahab
  Mirrokni}.} \bibinfo{year}{2017}\natexlab{}.
\newblock \showarticletitle{Budget management strategies in repeated auctions}.
  In \bibinfo{booktitle}{\emph{WWW}}. \bibinfo{publisher}{ACM},
  \bibinfo{address}{Perth, Australia}.
\newblock


\bibitem[Celli et~al\mbox{.}(2022)]%
        {celli2022parity}
\bibfield{author}{\bibinfo{person}{Andrea Celli}, \bibinfo{person}{Riccardo
  Colini-Baldeschi}, \bibinfo{person}{Christian Kroer}, {and}
  \bibinfo{person}{Eric Sodomka}.} \bibinfo{year}{2022}\natexlab{}.
\newblock \showarticletitle{The parity ray regularizer for pacing in auction
  markets}. In \bibinfo{booktitle}{\emph{WWW}}. \bibinfo{publisher}{ACM},
  \bibinfo{address}{Lyon, France}.
\newblock


\bibitem[Cuturi(2013)]%
        {cuturi2013sinkhorn}
\bibfield{author}{\bibinfo{person}{Marco Cuturi}.}
  \bibinfo{year}{2013}\natexlab{}.
\newblock \showarticletitle{Sinkhorn distances: Lightspeed computation of
  optimal transport}. In \bibinfo{booktitle}{\emph{NIPS}}.
  \bibinfo{publisher}{MIT Press}, \bibinfo{address}{Nevada, USA}.
\newblock


\bibitem[Edelman et~al\mbox{.}(2007)]%
        {edelman2007internet}
\bibfield{author}{\bibinfo{person}{Benjamin Edelman}, \bibinfo{person}{Michael
  Ostrovsky}, {and} \bibinfo{person}{Michael Schwarz}.}
  \bibinfo{year}{2007}\natexlab{}.
\newblock \showarticletitle{Internet advertising and the generalized
  second-price auction: Selling billions of dollars worth of keywords}.
\newblock \bibinfo{journal}{\emph{American economic review}}
  \bibinfo{volume}{97}, \bibinfo{number}{1} (\bibinfo{year}{2007}),
  \bibinfo{pages}{242--259}.
\newblock


\bibitem[Galichon(2018)]%
        {galichon2018optimal}
\bibfield{author}{\bibinfo{person}{Alfred Galichon}.}
  \bibinfo{year}{2018}\natexlab{}.
\newblock \bibinfo{booktitle}{\emph{Optimal transport methods in economics}}.
  Vol.~\bibinfo{volume}{1}.
\newblock \bibinfo{publisher}{Princeton University Press}.
\newblock


\bibitem[Li et~al\mbox{.}(2018)]%
        {li2018efficient}
\bibfield{author}{\bibinfo{person}{Pengcheng Li}, \bibinfo{person}{Ammar
  Hawbani}, {et~al\mbox{.}}} \bibinfo{year}{2018}\natexlab{}.
\newblock \showarticletitle{An efficient budget allocation algorithm for
  multi-channel advertising}. In \bibinfo{booktitle}{\emph{ICPR}}.
  \bibinfo{publisher}{IEEE}, \bibinfo{address}{Beijing, China}.
\newblock


\bibitem[Nemirovski(2004)]%
        {nemirovski2004interior}
\bibfield{author}{\bibinfo{person}{Arkadi Nemirovski}.}
  \bibinfo{year}{2004}\natexlab{}.
\newblock \showarticletitle{Interior point polynomial time methods in convex
  programming}.
\newblock \bibinfo{journal}{\emph{Lecture notes}} \bibinfo{volume}{42},
  \bibinfo{number}{16} (\bibinfo{year}{2004}), \bibinfo{pages}{3215--3224}.
\newblock


\bibitem[Nuara et~al\mbox{.}(2019)]%
        {nuara2019dealing}
\bibfield{author}{\bibinfo{person}{Alessandro Nuara}, \bibinfo{person}{Nicola
  Sosio}, \bibinfo{person}{Francesco Trov{\~A}}, \bibinfo{person}{Maria~Chiara
  Zaccardi}, \bibinfo{person}{Nicola Gatti}, {and} \bibinfo{person}{Marcello
  Restelli}.} \bibinfo{year}{2019}\natexlab{}.
\newblock \showarticletitle{Dealing with interdependencies and uncertainty in
  multi-channel advertising campaigns optimization}. In
  \bibinfo{booktitle}{\emph{WWW}}. \bibinfo{publisher}{ACM},
  \bibinfo{address}{San Francisco, USA}.
\newblock


\bibitem[Nuara et~al\mbox{.}(2018)]%
        {nuara2018combinatorial}
\bibfield{author}{\bibinfo{person}{Alessandro Nuara},
  \bibinfo{person}{Francesco Trovo}, \bibinfo{person}{Nicola Gatti}, {and}
  \bibinfo{person}{Marcello Restelli}.} \bibinfo{year}{2018}\natexlab{}.
\newblock \showarticletitle{A combinatorial-bandit algorithm for the online
  joint bid/budget optimization of pay-per-click advertising campaigns}. In
  \bibinfo{booktitle}{\emph{AAAI}}. \bibinfo{publisher}{The MIT Press},
  \bibinfo{address}{Louisiana USA}.
\newblock


\bibitem[Peyr{\'e} et~al\mbox{.}(2017)]%
        {peyre2017computational}
\bibfield{author}{\bibinfo{person}{Gabriel Peyr{\'e}}, \bibinfo{person}{Marco
  Cuturi}, {et~al\mbox{.}}} \bibinfo{year}{2017}\natexlab{}.
\newblock \showarticletitle{Computational optimal transport}.
\newblock \bibinfo{journal}{\emph{Center for Research in Economics and
  Statistics}} \bibinfo{volume}{1}, \bibinfo{number}{1} (\bibinfo{year}{2017}),
  \bibinfo{pages}{10--60}.
\newblock


\bibitem[Pisinger(1995)]%
        {pisinger1995minimal}
\bibfield{author}{\bibinfo{person}{David Pisinger}.}
  \bibinfo{year}{1995}\natexlab{}.
\newblock \showarticletitle{A minimal algorithm for the multiple-choice
  knapsack problem}.
\newblock \bibinfo{journal}{\emph{European Journal of Operational Research}}
  \bibinfo{volume}{83}, \bibinfo{number}{2} (\bibinfo{year}{1995}),
  \bibinfo{pages}{394--410}.
\newblock


\bibitem[Qin et~al\mbox{.}(2015)]%
        {qin2015sponsored}
\bibfield{author}{\bibinfo{person}{Tao Qin}, \bibinfo{person}{Wei Chen}, {and}
  \bibinfo{person}{Tie-Yan Liu}.} \bibinfo{year}{2015}\natexlab{}.
\newblock \showarticletitle{Sponsored search auctions: Recent advances and
  future directions}.
\newblock \bibinfo{journal}{\emph{ACM Transactions on Intelligent Systems and
  Technology}} \bibinfo{volume}{5}, \bibinfo{number}{4} (\bibinfo{year}{2015}),
  \bibinfo{pages}{1--34}.
\newblock


\bibitem[Santambrogio(2015)]%
        {santambrogio2015optimal}
\bibfield{author}{\bibinfo{person}{Filippo Santambrogio}.}
  \bibinfo{year}{2015}\natexlab{}.
\newblock \showarticletitle{Optimal transport for applied mathematicians}.
\newblock \bibinfo{journal}{\emph{Birk{\"a}user, NY}} \bibinfo{volume}{55},
  \bibinfo{number}{58-63} (\bibinfo{year}{2015}), \bibinfo{pages}{94}.
\newblock


\bibitem[Sinha and Zoltners(1979)]%
        {sinha1979multiple}
\bibfield{author}{\bibinfo{person}{Prabhakant Sinha} {and}
  \bibinfo{person}{Andris~A Zoltners}.} \bibinfo{year}{1979}\natexlab{}.
\newblock \showarticletitle{The multiple-choice knapsack problem}.
\newblock \bibinfo{journal}{\emph{Operations Research}} \bibinfo{volume}{27},
  \bibinfo{number}{3} (\bibinfo{year}{1979}), \bibinfo{pages}{503--515}.
\newblock


\bibitem[Villani(2009)]%
        {villani2009optimal}
\bibfield{author}{\bibinfo{person}{C{\'e}dric Villani}.}
  \bibinfo{year}{2009}\natexlab{}.
\newblock \bibinfo{booktitle}{\emph{Optimal transport: old and new}}.
  Vol.~\bibinfo{volume}{338}.
\newblock \bibinfo{publisher}{Springer}.
\newblock


\bibitem[Wei et~al\mbox{.}(2019)]%
        {wei2019optimal}
\bibfield{author}{\bibinfo{person}{Chao Wei}, \bibinfo{person}{Weiru Zhang},
  \bibinfo{person}{Shengjie Sun}, \bibinfo{person}{Fei Li},
  \bibinfo{person}{Xiaonan Meng}, \bibinfo{person}{Yi Hu},
  \bibinfo{person}{Kuang-chih Lee}, {and} \bibinfo{person}{Hao Wang}.}
  \bibinfo{year}{2019}\natexlab{}.
\newblock \showarticletitle{Optimal Delivery with Budget Constraint in
  E-Commerce Advertising}. In \bibinfo{booktitle}{\emph{RecSys}}.
  \bibinfo{publisher}{PMLR}, \bibinfo{address}{Peter Knees, TU Wien}.
\newblock


\bibitem[Wilkens et~al\mbox{.}(2017)]%
        {wilkens2017gsp}
\bibfield{author}{\bibinfo{person}{Christopher~A Wilkens},
  \bibinfo{person}{Ruggiero Cavallo}, {and} \bibinfo{person}{Rad Niazadeh}.}
  \bibinfo{year}{2017}\natexlab{}.
\newblock \showarticletitle{GSP: the cinderella of mechanism design}. In
  \bibinfo{booktitle}{\emph{WWW}}. \bibinfo{publisher}{ACM},
  \bibinfo{address}{Perth, Australia}.
\newblock


\bibitem[Xu et~al\mbox{.}(2015)]%
        {xu2015smart}
\bibfield{author}{\bibinfo{person}{Jian Xu}, \bibinfo{person}{Kuang-chih Lee},
  \bibinfo{person}{Wentong Li}, \bibinfo{person}{Hang Qi}, {and}
  \bibinfo{person}{Quan Lu}.} \bibinfo{year}{2015}\natexlab{}.
\newblock \showarticletitle{Smart pacing for effective online ad campaign
  optimization}. In \bibinfo{booktitle}{\emph{SIGKDD}}.
  \bibinfo{publisher}{ACM}, \bibinfo{address}{Sydney, Australia}.
\newblock


\bibitem[Zhang et~al\mbox{.}(2021)]%
        {zhang2021optimizing}
\bibfield{author}{\bibinfo{person}{Zhilin Zhang}, \bibinfo{person}{Xiangyu
  Liu}, \bibinfo{person}{Zhenzhe Zheng}, \bibinfo{person}{Chenrui Zhang},
  \bibinfo{person}{Miao Xu}, \bibinfo{person}{Junwei Pan},
  \bibinfo{person}{Chuan Yu}, \bibinfo{person}{Fan Wu}, \bibinfo{person}{Jian
  Xu}, {and} \bibinfo{person}{Kun Gai}.} \bibinfo{year}{2021}\natexlab{}.
\newblock \showarticletitle{Optimizing Multiple Performance Metrics with Deep
  GSP Auctions for E-commerce Advertising}. In
  \bibinfo{booktitle}{\emph{WSDM}}. \bibinfo{publisher}{ACM},
  \bibinfo{address}{Jerusalem, Israel}.
\newblock


\bibitem[Zhao et~al\mbox{.}(2019)]%
        {zhao2019unified}
\bibfield{author}{\bibinfo{person}{Kui Zhao}, \bibinfo{person}{Junhao Hua},
  \bibinfo{person}{Ling Yan}, \bibinfo{person}{Qi Zhang}, \bibinfo{person}{Huan
  Xu}, {and} \bibinfo{person}{Cheng Yang}.} \bibinfo{year}{2019}\natexlab{}.
\newblock \showarticletitle{A Unified Framework for Marketing Budget
  Allocation}. In \bibinfo{booktitle}{\emph{SIGKDD}}. \bibinfo{publisher}{ACM},
  \bibinfo{address}{Anchorage, Alaska, USA}.
\newblock


\bibitem[Zhou et~al\mbox{.}(2018)]%
        {zhou2018deep}
\bibfield{author}{\bibinfo{person}{Guorui Zhou}, \bibinfo{person}{Xiaoqiang
  Zhu}, \bibinfo{person}{Chenru Song}, \bibinfo{person}{Ying Fan},
  \bibinfo{person}{Han Zhu}, \bibinfo{person}{Xiao Ma},
  \bibinfo{person}{Yanghui Yan}, \bibinfo{person}{Junqi Jin},
  \bibinfo{person}{Han Li}, {and} \bibinfo{person}{Kun Gai}.}
  \bibinfo{year}{2018}\natexlab{}.
\newblock \showarticletitle{Deep interest network for click-through rate
  prediction}. In \bibinfo{booktitle}{\emph{SIGKDD}}. \bibinfo{publisher}{ACM}.
\newblock


\bibitem[Zhu et~al\mbox{.}(2017)]%
        {zhu2017optimized}
\bibfield{author}{\bibinfo{person}{Han Zhu}, \bibinfo{person}{Junqi Jin},
  \bibinfo{person}{Chang Tan}, \bibinfo{person}{Fei Pan},
  \bibinfo{person}{Yifan Zeng}, \bibinfo{person}{Han Li}, {and}
  \bibinfo{person}{Kun Gai}.} \bibinfo{year}{2017}\natexlab{}.
\newblock \showarticletitle{Optimized cost per click in taobao display
  advertising}. In \bibinfo{booktitle}{\emph{SIGKDD}}.
  \bibinfo{publisher}{ACM}.
\newblock


\end{thebibliography}

\end{document}